\documentclass{emulateapj}
\usepackage{natbib}
\usepackage{xspace}
\usepackage{amssymb}
\usepackage{amsmath}
\usepackage{graphicx}
\usepackage{epstopdf}

\shorttitle{Determining Distances to Clusters of Galaxies using 
Resonant X-ray Lines}
\shortauthors{S.~M.~Molnar, M.~Birkinshaw and R.~F.~Mushotzky}

\newcommand{\simless} {\ensuremath{\lower 
3pt\hbox{$\rlap{\raise5pt\hbox{$\char'074$}}\mathchar"7218$}}}
\newcommand{\simgreat} {\ensuremath{\lower 
3pt\hbox{$\rlap{\raise5pt\hbox{$\char'076$}}\mathchar"7218$}}}

\begin{document}

\title{Determining Distances to Clusters of Galaxies using Resonant 
X-ray Emission Lines}

\author{
Sandor M. Molnar\altaffilmark{1}, Mark Birkinshaw\altaffilmark{2} 
                    and Richard F. Mushotzky\altaffilmark{3}
}

\begin{abstract} 
Bright clusters of galaxies can be seen out to cosmological distances, 
and thus they can be used to derive cosmological parameters.
Although the continuum X-ray emission from the intra-cluster gas is 
optically thin, the optical depth of resonant lines of ions of heavy 
elements can be larger than unity. 
In this {\it Letter} we study the feasibility of deriving
distances to clusters of galaxies by determining the spatial 
distribution 
of the intra-cluster gas from X-ray imaging and the optical depth from 
resonant emission lines (the XREL method).
We solve the radiative transfer problem for line scattering in the hot
intra-cluster gas using Monte Carlo simulations.
We discuss the spatial and spectral resolutions
needed to use the XREL method for accurate determination of distances,
and hence cosmological parameters, and show that accurate
distances will be obtained by applying this technique with the next
generation of high resolution X-ray spectrometers.
\end{abstract}


\altaffiltext{1}{Institute for Theoretical Physics, University of 
Zurich,
                  190 Winterthurersgtrasse, Zurich, CH-8057, 
Switzerland}
\altaffiltext{2}{Department of Physics and Astronomy, University of 
Bristol,
                  Tyndall Avenue, Bristol, BS8 1TL, UK}
\altaffiltext{3}{Laboratory for High Energy Astrophysics, NASA/Goddard 
Space Flight
                 Center, Greenbelt, MD 20771, USA}




\section{Introduction}
\label{S:Intro}

Direct physical distance measurements are of paramount importance in 
cosmology, since they provide us with a reliable distance
scale, and make it possible to confirm or rule out theoretical
cosmological models. Direct distance measurements are free from
the systematic errors inherent in the cosmic distance ladder method
\citep{Web99}. Clusters of galaxies, the known largest
gravitationally bound structures in the Universe, have been used 
extensively to determine cosmological parameters taking advantage of 
the fact that their formation and evolution are sensitive
to the underlying cosmology (Allen et al. 2004; Patrick 2004; 
Ettori, Tozzi and Rosati 2003; Vikhlinin et al.
2003; Viana et al. 2003; Holder, Haiman and Mohr 2001; 
Haiman, Mohr and Holder, 2001; 
for a review see Rosati, Borgani and Norman 2002).
Clusters of galaxies can also be used as ``standard candles'' to 
determine cosmological parameters via distance measurements in the 
same way as the supernova method (Bonamente et al. 2005; Carlstrom, 
Holder and Reese, 2002; Mason, Myers, and Readhead 2001; 
Pointecouteau et~al. 2001; Komatsu et~al. 1999; for future prospects 
see Molnar, Birkinshaw and Mushotzky, 2002; Molnar et al. 2004). 
The next generation of such programmes aim to
determine cosmological parameters to a statistical accuracy of a few
percent, and hence are likely to be limited by systematic errors 
\citep{Molnet04,Perl03,Haiet01,Holet01}
New distance measurement techniques are valuable because they permit 
cross-checks of measurements of cosmological parameters made using 
completely independent methods.

In this {\it Letter} we propose to exploit a new technique, 
the X-ray emission line (XREL) method, to determine direct 
distances to any object which, potentially, has optical depths 
in x-ray resonance lines on the order of 1-5. 
This method is similar to the Sunyaev-Zel'dovich -- X-ray (SZX)
method (Sunyaev and Zel'dovich, 1980; for recent 
reviews see Carlstrom, Holder and Reese, 2002; Birkinshaw, 1999), 
and to the X-ray resonant line absorption method, which
is based on an absorption feature in resonance 
lines in spectra of bright AGNs behind the cluster 
\citep{Sar89,KroRay88}.
We discuss the applicability of this technique to massive clusters of 
galaxies via Monte Carlo simulation of radiative transfer in the 
Fe He-like resonant line.




\section{X-ray Emission Lines}
\label{S:Lines}

X-ray spectral lines contain a plethora of information about the 
physical and chemical conditions of the intra-cluster gas 
\citep{Mush03}.
As has been pointed out by \citet*{Gilf87},
resonant lines of ions of heavy elements can have optical depths
larger than unity in clusters. Photons in these
lines suffer multiple scattering, and the line shapes and
intensities change as a result \citep{Gilf87,Shig98}, so that 
the line amplitudes are reduced at the core of the cluster and
enhanced in the outer regions. Ignoring this effect can cause an
underestimate in the  
abundances based on resonance lines \citep{Shig98,Akim00}.
Scatterings can also polarize the emission lines \citep*{Sazo02}.

The line shape is also affected by gas motion in clusters.
This can reduce the optical depth of all lines by Doppler 
broadening \citep{Chu04}. 
In the center of Virgo cluster (M87), 
Mathews, Buote and Brighenti (2001) estimated that 
turbulent motion with a Mach number of 0.3 can reduce the
optical depth substantially (from 2.5 to 1 in the case of the Si line).
Sakelliou et al. (2002), using XMM-{\it Newton} observations, 
found no evidence for resonant scattering in M87.
On the other hand Xu et al. (2001) found evidence for resonant line 
scattering at NGC 4636 in the outskirts of the Virgo cluster.
In the Perseus cluster resonant scattering was ruled out based on an
analysis of the Fe XXVI K lines (Gastaldello and Molendi 2004;
Churazov et al. 2004).  
However, Sanders, Fabian and Dunn (2005) argue that the existence of 
H$\alpha$ filaments in the core of Perseus (Conselice, Gallagher and
Wyse 2001) suggests no turbulence. Based on resonance lines of 
Fe XXIV and Fe XXIII, Sanders et al. (2005) conclude that, in Perseus, 
resonance scattering may be the cause of the apparent central drop in 
metal abundances. Clearly, more observations are needed to 
clarify the role of turbulence in clusters.
The effects of turbulence and resonant scattering can be separated, 
in principle, since scattering broadens only resonant 
lines with non-negligible optical depth, while turbulence broadens all 
lines. 
Mild turbulence (up to Mach numbers of a few tenths) may leave
some lines with optical depth larger than unity, so that it remains
possible to determine the optical depth from the line shapes.
In such cases we could use the shapes of optically-thin lines as a 
template
and calculate the broadening due to resonance line scattering
\citep[see Figures 1 and 2 in][]{InogSuny03}.




\section{The X-ray -- Resonance Emision Line Method}
\label{S:Method}

The XREL method compares the
central X-ray surface brightness of a cluster generated by
thermal bremsstrahlung, $S_{X0} \propto n_{e0}^2 \,
\Lambda_e(T_e,Z_{ab}) \, L \, (1 + z)^{-4}$, where $n_{e0}$ is
the central electron number density, $\Lambda_e$ is the X-ray
emissivity as a function of electron temperature and
heavy element abundance, and $L$ is a line-of-sight size of
the cluster, with a quantity linear in $n_{e0} \, L$. In the SZX
method that quantity is the central SZ effect or a measure
proportional to it. In the XREL method we use the optical depth
from the center of the cluster to the observer in the chosen
resonant emission line, $\tau_0 \propto n_{e0} L$. The ratio
$\tau_{0}^2/S_{X0}$ then provides a measure of $L$ which can be
compared with the angular size of the cluster to estimate the
angular diameter distance, $D_A$, at the cluster redshift $z$.

The determination of $D_A$ in this way requires an understanding
of the shape of the cluster, and this is determined from the
the X-ray image. In this {\it Letter} we study the main
features of resonant line scattering using a spherical isothermal
beta model for the cluster gas \citep{CaFu78}.
A later paper will study
scattering in atmospheres drawn from numerical simulations, to
investigate the effect of more realistic 3D cluster shapes and
cluster evolution.
Assuming an isothermal beta model, $D_A$ can be expressed as
\begin{equation}
  D_A \propto \bigl( 1 + z \bigr)^{-4} \; F(\theta_c,\beta) \;
              \Lambda_e(T_e, Z_{ab}) \; \tau_0^2 \; S_{X0}^{-1}
  ,  
  \label{E; DA}
\end{equation}
where $F(\theta_c,\beta)$ is a form factor that depends on the 
angular core radius of the cluster, $\theta_c = R_c/D_A$, and 
the shape parameter $\beta$. Once the angular diameter distance --
redshift function, $D_A(z)$, is determined by applying
Equation~\ref{E; DA} for several clusters, it can be used to
constrain cosmological parameters \citep{Molnet02}.

If the sensitivity and the spatial resolution of the
instrument are high enough, multiple distances to a cluster can be
obtained by studying several resonant lines in different regions 
in the image of the cluster.
The self checking that this provides can test for systematic errors 
in the XREL method. Determination of optical depths using
several resonance lines and/or line intensities could provide 
additional 
tests.

In the case of non-spherical clusters, the XREL
method is subject to the same limitations as the SZX method, in
particular to biases involved in the comparison of line-of-sight and
angular extents of clusters. 
Thus an alternative application of the XREL method might be
to help in de-projecting cluster density and temperature
distributions if a cluster is at a well-known distance and 
suspected to be non-spherical.




\section{Determination of Optical Depth}
\label{S:TAU}

The optical depth in a resonant emission line can, in principle, be
determined from the line profile, from the spatial distribution of
the line's intensity, or from the polarization of the lines, and 
used to determine distances to clusters of galaxies
(Sazonov et al. 2002; Sarazin 1989).
In this {\it Letter} we focus on the determination of the optical 
depth based on the line profile.

The most prominent resonant line in hot clusters of galaxies 
is that of He-like iron (Fe~XXV) at 6.7 keV, which 
can be detected at cosmological distances. 
The optical depth of this line can be larger than unity,
although there are significant variations in the estimates 
in the literature (compare, for example Krolik and Raymond 1988 and
Sarazin 1989).
The spatial distribution of the emitted photons in the rest frame of 
the ion is a mixture of an isotropic and a dipole pattern with 
weights depending on the change in the total angular momentum between 
the two energy states participating in the transition.

We have solved the radiative transfer problem of line scattering in
the intra-cluster gas via Monte Carlo simulations.
This problem is well suited to Monte Carlo simulation
since the optical depth in the resonant lines is not excessive. 
We determined the line profile of the 6.7 keV Fe line emerging from
clusters of galaxies for spherically-symmetric isothermal beta models
for a range of optical depths, $\tau_0$, from 0.1 to~5. 
We carried out simulations using different values of $\beta$
within the range $0.6 - 0.9$ suggested by cluster data.
We found that the line profile emerging from the core
of a cluster is sensitive to $\tau_0$, not to $\beta$ 
(differs $\le$ 1\%, for fixed $\tau_o$), thus, for our purposes 
in this {\it Letter}, we present results for only $\beta = 2/3$.
We also included the natural line width (FWHM = 0.335 eV), 
which is about 10\% of the thermal  line broadening for hot (5 - 10 keV) 
clusters of galaxies.
We assumed two temperatures 5 and 10 kev for our simulations, 
and we worked with frequency shifts normalized by the Doppler thermal
width, $\Delta x = (\nu - \nu_0)/\Delta \nu_D$, and 
lengths normalized to the cluster core radius.
Since the 6.7 keV Fe line originates from a pure dipole transition, 
there is a correlation between photon propagation and the
polarization, and so the polarization character of the emission 
lines was included in our code. Further details of the code can be
found in \citet{Molnet99}. 
In the high temperature, low density, intra-cluster gas, 
collision times are much longer than the spontaneous emission times. 
Thus, to an excellent level of approximation, we may assume that 
photons once emitted, are not destroyed, but scatter
and escape from the cluster.

\begin{figure}
\centerline{
\includegraphics[width=7.2cm]{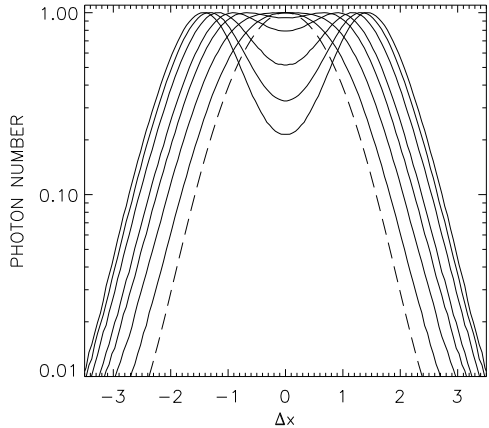}
}
\caption{
 Emerging line profiles from spherical isothermal beta
 models at a range of optical depths. See text for details. 
\label{F:TAUALL}
}
\end{figure} 

In Figure~\ref{F:TAUALL} we show the emergent line profiles for
different optical depths, $\tau_0$ = 0.5, 1, 1.5, 2, 3, 4, 5 
(5 is the broadest) for the integrated X-ray emission from the 
center of a cluster with impact parameters, $d$, 
less than half the core radius ($d < 0.5 \, R_c$).
The dashed line represents the thermally-broadened Gaussian line profile
characteristic of optically-thin emission. This figure demonstrates
the diffusion of photons in frequency space. 
As a result of scattering, the line broadens. 
Photons can escape easily from the wings of the line where the optical 
depth is low, but the line center will be depleted. 
The number of photons at the wings of the line is further
increased by photons scattered into the line of sight from 
outside the nominal field of view.
As a result a double-peaked line profile arises in the cores of 
clusters with optical depths $\simgreat \, 2$, which is a 
unique signature of multiple scatterings. 
Comparing double-peaked lines and optically thin lines
makes it possible to separate resonant scattering and turbulence.




\section{Determination of Cosmological Parameters}
\label{S:CosParam}

In principle the angular diameter distance redshift function, 
$D_A(z)$, is most sensitive to $\Omega_\Lambda$ at a redshift
of around unity. 
Due to observational constraints, the upper range of
the ideal redshift window will be somewhat less than one. 
Therefore, measurements of $D_A(z)$ for a sample of clusters of
galaxies distributed in the redshift range of zero to about one can 
be expected to provide useful constraints on the Hubble
parameter ($h = H_0 / 100$ km s$^{-1}$ Mpc$^{-1}$) and the energy
densities, ($\Omega_m, \Omega_\Lambda)$, or the equation of
state parameter, $w = p/\rho$. The accuracy to which 
we can determine these cosmological
parameters depends on the accuracy of, and the degeneracies in, the
$D_A(z)$ function (a detailed study of 
errors, their degeneracies and their effect on the errors in the 
cosmological parameters is given in Molnar et al.~2002).

\begin{figure}
\centerline{
\includegraphics[width=7.2cm]{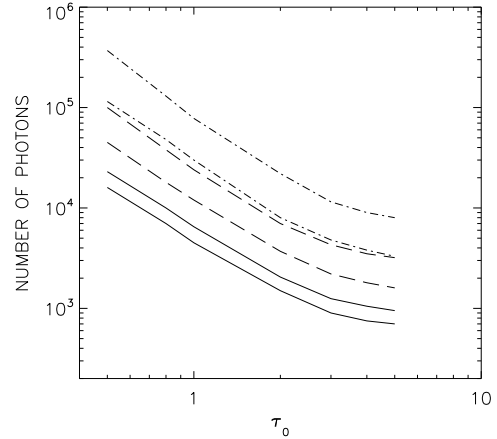}
}
\caption{
 The number of photons necessary to obtain 5\% accuracy (1 $\sigma$)
 in $\tau_0$ for spherical isothermal beta models with 
 $T_X = 5$~or 10~keV using spectrometers with 2~eV (solid), 4~eV
 (dashed), or 6~eV (dot-dashed) resolution. See text for details.
} \label{F:NPHOT}
\end{figure} 

Errors in the angular diameter distance from the XREL method will be 
dominated by uncertainties in $F(\theta_c,\beta)$, $S_{X0}$, and 
$\tau_0$, errors from the other terms will be negligible by comparison. 

The accuracy of $\tau_0$ crucially depends on 
the spatial and spectral resolution of the instrument used.
The amplitude of the dip at line center of a double-peaked line is 
sensitive to $\tau_0$, and so for the best accuracy this central 
dip should be spectrally resolved. 
An instrument with an angular resolution $\sim \theta_c/2$ looking at
the cluster core will observe a line profile close to that seen by an
instrument with high spatial resolution. Therefore, high spatial
resolution is not necessary to determine $\tau_0$.
On the other hand high spectral resolution is crucial.

We carried out Monte Carlo simulations to study the feasibility of
accurate optical depth determination based on the Fe 6.7 keV resonant 
line. We derived the number of photons in the line necessary to determine 
$\tau_0$, with a fixed accuracy assuming spectroscopically-determined 
intra-cluster gas temperatures $T_X$ 
(assumed to be an accurate representation of the electron temperature, 
$T_e$) of 5 and 10~keV, which is a range for massive, observable high 
redshift clusters.
We collected photons with an impact parameter less than half the core 
radius ($d < 0.5 \, R_c$). 
We assumed an Fe abundance of 0.3 in units of solar abundance, 
and a cluster X-ray luminosity - temperature scaling \citep{Willet05}.
Figure~\ref{F:NPHOT} shows the number of photons in the line required
to determine $\tau_0$ to 5\% accuracy (CL 68\%). 
Solid, dashed and dash-dotted lines represent X-ray
observations carried out using instruments with a spectral
resolution of 2, 4 and 6 eV (FWHM). The upper (lower) lines represent 
results for $T_X = 5$ keV ($T_X = 10$ keV). 
The average frequency change per scattering is larger for
high temperature clusters, thus we need fewer photons to
determine accurate optical depths for these clusters.
Clearly, high-accuracy optical depth determination is feasible only for 
$\tau_0 \, \simgreat \, 2$, in which case we would need a few thousand 
photons in the Fe 6.7 keV resonant line, assuming a spectral resolution 
of 2 - 4 eV. A lower-resolution spectrometer of $6$~eV FWHM 
would require about ten times more line photons than
a spectrometer with an FWHM of 2~eV.

Practical application of this method is likely to encounter a
5\% statistical error in $F(\theta_c,\beta)$, 
and a similar error in $S_{X0}$ \citep{Molnet02}.  
In addition, assuming a 5\% statistical error in the optical depth, 
if the cluster shape is well determined, the angular diameter 
distance can be determined with 10\% accuracy.
Measuring the angular diameter distance to only 100
clusters in the redshift interval $0 \le z \le 1$ with a 10\% 
statistical
error in $D_A(z)$, and assuming an $\Omega_m + \Omega_\Lambda = 1$
(spatially flat) prior, would allow us to determine $\Omega_m$ and
$\Omega_\Lambda$, with 16\% and 10\% (3$\,\sigma$) errors.
Adding an SZ survey would lead to a measurement of $w$ 
with a 20\% (3$\,\sigma$) error \citep{Molnet04}. 
This is comparable to the achievable accuracy based on the supernova 
method: using 63 supernovae, assuming a spatially flat model and 
including {\it WMAP} constraints, $\Omega_m$, $\Omega_\Lambda$ and 
$w$ can be determined to about 30\%, 30\% and 15\% (1$\,\sigma$) 
accuracy \citep{Knoet03}.

With the next generation of X-ray instruments, it should be possible
to obtain spectra with a few thousand photons in the Fe 6.7 keV line. 
Even today's instruments (XMM-{\it Newton}) can collect
10$^5$ photons in 50 ks in the Fe 6.7 keV line complex
from the core of Perseus cluster \citep{InogSuny03}.
Future missions 
(Constellation-X\footnote{http://constellation.gsfc.nasa.gov},
XEUS\footnote{http://www.rssd.esa.int/index.php?project=XEUS}) will 
have ten times more effective area and about 4~eV spectral resolution.
\citet{Bonaet05} used {\it Chandra} observations of 39 clusters 
(with a total of ~6000 ks of data) and SZ observations to
determine the Hubble constant with a 4.5\% statistical error.
Using XMS on Constellation-X, it will be possible to obtain about 2000 
photons in the 6.7 Fe K$\alpha$ resonant line within half a core radius 
with a 15 ks exposure time and determine the optical depth of clusters 
located at about $z = 0.1$ with a 5\% error.
Taking account of uncertainties in the cluster shape, this should lead
to a distance error $\sim 20$\%, so that observations of a
sample of 30 clusters for a total exposure time of 450 ks should 
determine the Hubble constant with about a 3\% error.

High redshift clusters need much longer exposures to obtain
a few thousand photons in the 6.7 Fe resonant line within half a core
radius of the cluster: to get useful coverage to $z \sim 0.8$ with
Constellation-X an exposure time of about 1500 ks would be
needed per cluster. 
However, this is likely to be an overestimate of the exposure
time --- double the number of photons per cluster can be used if the
spatial region is increased, and for many clusters it should be possible 
to include other resonant lines (for example from Si and S).
Also, cold core clusters have much higher Fe abundances
at the center than we assumed in our simulations.
Massive relaxed clusters with $T_X \ge 8$ keV such as 
A1689 (z=0.183), ZW3146 (z=0.291), MS 0451-03 (z = 0.54), and
MS 1054-03 (z=0.833) seem to be good candidates.                                                                              
A detailed feasibility study is needed to determine the best combined
spatial/spectral strategy and the optimal redshift distribution of the 
cluster sample, but this is beyond the scope of the present {\it 
Letter}.




\section{Conclusion}
\label{S:Conclusion}

In this {\it Letter} we have estimated the achievable accuracy in
cosmological parameters based on direct distance measurements to
clusters of galaxies determined from X-ray imaging and resonant
emission lines (the XREL method). 
Our feasibility study concludes that a moderate spatial resolution
(about a half of the core radius, or $30 - 60$~arcsec at the 
relevant redshifts) and high spectral resolution (ideally about 2 eV) 
are necessary to resolve the central dip in the Fe XXV 6.7~keV line
produced by optical depth effects, and thus to achieve a useful, 10\%
cosmological distance accuracy. 
The next generation X-ray satellites will have the necessary 
sensitivity and resolution to apply the XREL method in this way, and 
so to provide a cross-check on other methods of measuring cosmological 
parameters accurately.

\acknowledgments
We thank G. Lake, J. Stadel and Z. Haiman 
for comments on our paper, P. Saha for discussions, and
the referee, S. Ettori, for suggestions which helped to clarify 
certain points of our paper.

%
%
\bibliographystyle{apj}


 
\end{document}